\documentclass[namedreferences]{solarphysics}

\usepackage[hyperref,optionalrh,showbiblabels]{spr-sola-addons} 
\usepackage{graphicx}                                                                     
\usepackage{amssymb}                                                                   
\usepackage{color}                                                                           
\usepackage{breakurl}                                                                      

\renewcommand{\vec}[1]{{\mathbfit #1}}

\newcommand{\curl}{ {\bf \nabla} \times}

\newcommand{\arcsec}{^{\prime\prime}}

\newcommand{\bb}{\vec B}

\newcommand{\aap}{    {\it Astron. Astrophys.}}

\newcommand{\apj}{    {\it Astrophys. J.}}
\newcommand{\apjl}{   {\it Astrophys. J. Lett.}}
\newcommand{\apss}{   {\it Astrophys. Space Sci.}}

\newcommand{\mnras}{  {\it Mon. Not. Roy. Astron. Soc.}}
\newcommand{\nat}{    {\it Nature}}

\newcommand{\pasj}{   {\it Pub. Astron. Soc. Japan}}

\newcommand{\raa}{    {\it Research in Astronomy and Astrophysics}}
\newcommand{\sci}{    {\it Science}}
\newcommand{\solphys}{{\it Solar Phys.}}
 
\newcommand{\ssr}{    {\it Space Sci. Rev.}} 
\chardef\us=`\_

\begin{document}
\begin{article}
\begin{opening}

\title{Observations of multiple blobs in homologous solar coronal jets in closed loops\\ {\it Solar Physics}}

\author{Q.~M.~\surname{Zhang}$^{1}$\sep
            H.~S.~\surname{Ji}$^{1}$\sep
            Y.~N.~\surname{Su}$^{1}$
            }
 
\institute{$^{1}$ Key Laboratory for Dark Matter and Space Science, Purple Mountain Observatory, CAS, Nanjing 210008, China
                  email: \url{zhangqm@pmo.ac.cn}, email: \url{jihs@pmo.ac.cn}, email: \url{ynsu@pmo.ac.cn} \\
             }

\runningauthor{Zhang et al.}
\runningtitle{Multiple blobs in homologous coronal jets}

\begin{abstract}
Coronal bright points (CBPs) and jets are ubiquitous small-scale brightenings that are often associated with each other. In this paper, we report our 
multiwavelength observations of two groups of homologous jets. The first group was observed by the \textit{Extreme-Ultraviolet Imager} (EUVI) aboard 
the behind \textit{Solar TErrestrial RElations Observatory} (STEREO) spacecraft in 171 {\AA} and 304 {\AA} on 2014 September 10, from a location where 
data from the \textit{Solar Dynamic Observatory} (SDO) could not observe. The jets (J1$-$J6) recurred for six times with 
intervals of 5$-$15 minutes. They originated from the same primary CBP (BP1) and propagated in the northeast direction along large-scale, closed coronal loops.  
Two of the jets (J3 and J6) produced sympathetic CBPs (BP2 and BP3) after reaching the remote footpoints 
of the loops. The time delays between the peak times of BP1 and BP2 (BP3) are 240$\pm$75 s (300$\pm75$ s). The jets were not coherent. Instead, 
they were composed of bright and 
compact blobs. The sizes and apparent velocities of the blobs are 4.5$-$9 Mm and 140$-$380 km s$^{-1}$, respectively.
The arrival times of the multiple blobs in the jets at the far-end of the loops indicate that the sympathetic CBPs are caused by jet flows 
rather than thermal conduction fronts. The second group was observed by the \textit{Atmospheric Imaging Assembly} aboard SDO in various 
wavelengths on 2010 August 3. Similar to the first group, the jets originated from a short-lived bright point (BP) at the boundary of active region 11092 
and propagated along a small-scale, closed loop before flowing into the active region. Several tiny blobs with sizes of $\sim$1.7 Mm and apparent velocity 
of $\sim$238 km s$^{-1}$ were identified in the jets. We carried out the differential emission measure (DEM) inversions to investigate the temperatures of 
the blobs, finding that the blobs were multithermal with average temperature of 1.8$-$3.1 MK. The estimated number densities of the blobs were 
(1.7$-$2.8)$\times$10$^9$ cm$^{-3}$.
\end{abstract}
\end{opening}

\section{Introduction} \label{s-intro}
Coronal jets are transient and collimated plasma motions accompanied by point-like or loop-like brightenings at their bottom. They were first discovered in soft X-ray (SXR) 
by the YOHKOH spacecraft observations \citep{shi92}. With the development of solar space telescopes and the increase of spatial resolution, more and more 
coronal jets are observed and investigated in SXR as well as extreme-ultraviolet (EUV) wavelengths \citep{shi96,shi00,para15,nis15}. They are located not only in coronal 
holes and active region (AR) boundary where open magnetic field dominates, but also in quiet regions with large-scale, closed magnetic field \citep{cir07,cul07,zhang12b}. The 
typical length of jets is 10$-$400 Mm, the width is 5$-$100 Mm, and the apparent velocity is 10$-$1000 km s$^{-1}$ \citep{shi96}. The temperature of polar jets ranges from 0.1 
to 6.2 MK, with an average value of 1.8 MK. The electron number density of polar jets ranges from 0.1 to 8.0$\times$10$^8$ cm$^{-3}$, with an average value of 1.5$\times$10$^8$ 
cm$^{-3}$ \citep{para15}. These parameters for polar jets are much lower than those for flare-related jets where much larger amount of free energies are released \citep{shi00}.
It is generally accepted that coronal jets are caused by magnetic reconnection. The way of reconnection, however, depends largely on the magnetic configuration. 
In the two-dimensional (2D) case, new magnetic fluxes emerge from beneath the photosphere and reconnect with the pre-existing, open magnetic fields of opposite polarity, 
resulting in hot, collimated jets and two bright lobes at the base of jets \citep{hey77,shi92,yoko96}. Such mechanism became popular and was developed in the state-of-the-art, 
three-dimension (3D) magnetohydrodynamic (MHD) numerical experiments \citep{more08,more13,arch13,fang14}. A fraction of rotating coronal jets might be produced by the 
magnetic reconnection as a result of swirling motion of an embedded bipole in the photosphere \citep{par09,par10,wyp15}. \citet{moo10} classified coronal jets into the standard type 
and blowout type. The standard type with simpler morphology can be explained by the emerging-flux model \citep{shi92,lim15}. The blowout type usually results from small-scale filament 
eruptions \citep{moo13}. They have cool components and show rotating and/or transverse drifting motions \citep{chen12,hong13,puc13,sch13,zhang14a,liu15b}. However, \citet{ster15} 
proposed that both the standard and blowout polar jets in their sample are driven by small-scale filament eruptions, which resemble the coronal mass ejections (CMEs) driven by 
typical filament eruptions 
\citep{lin04}. Sometimes, a jet recurs at the same place with the same morphology and direction of outflow, forming the so-called recurrent or homologous jets. The recurrent jets 
may result from recurring magnetic reconnection \citep{chi08,inn11,wang12,zhang12b,guo13,chand15}, moving magnetic features \citep{yang13}, continuous twisting 
motion of the photosphere \citep{par10}, emergence of a bipole below transequatoral loop \citep{jia13}, or repeated cancellation of the pre-existing magnetic flux by the newly 
emerging flux \citep{chae99,chen15,li15}.

For the first time, \citet{zhang14b} reported the discovery of recurring blobs in homologous EUV jets observed by the \textit{Atmospheric Imaging Assembly} \citep[AIA;][]{lem12} 
aboard the \textit{Solar Dynamics Observatory} (SDO) with unprecedented temporal and spatial resolutions. Using six of the optically-thin filters (94, 131, 171, 193, 211, 335 {\AA}) 
of AIA, the authors performed the Differential Emission Measure (DEM) inversion and derived the DEM profiles of the recurring blobs as a function of temperature. It is found that 
the bright and compact features with average size of $\sim$3 Mm are multithermal in nature with average temperature of $\sim$2.3 MK. They are ejected outwards along the jet flow 
at speeds of 120$-$450 km s$^{-1}$. Most of the blobs have lifetime of 24$-$60 s before merging with the background plasma and disappearing. Such intermittent magnetic plasmoids 
or blobs created by magnetic reconnection have also been observed in the large-scale current sheet (CS) driven by CMEs and the small-scale CS associated with chromospheric jets 
\citep{asai04,lin05,tak12,sin12,kum13,lin15}. They are generally explained by the tearing-mode instability (TMI) of the thin CS where a series of magnetic islands are recurrently created 
during magnetic reconnection \citep{fur63,dra06,bar08,inn15}, which makes electron acceleration more efficient and dynamic \citep{kli00}. Recently, multidimensional, 
MHD numerical simulations of TMI have been significantly improved, thanks to the rapid advancement of powerful supercomputers. \citet{yang13} performed 2.5-dimensional 
(2.5D; $\partial/\partial z=0$) numerical experiments to simulate the process of magnetic reconnection between the moving magnetic features and the pre-existing magnetic 
field. The experiments successfully reproduce the plasmoids as a result of TMI, which are consistent with the observed bright moving blobs in the chromospheric anemone jets.
The plasmoids have temperatures of $\sim$0.015 MK, densities of $\sim$1.5$\times$10$^{14}$ cm$^{-3}$, and sizes of 0.05$-$0.15 Mm. They move bidirectionally at speeds 
of $\sim$30 km s$^{-1}$, which is close to the local Alfv\'{e}n speed. After reaching the outflow regions, the plasmoids collide with the magnetic fields there and are quickly destroyed 
as they disappear. Using a Harris CS, \citet{ni15} performed 2.5D numerical simulations of magnetic reconnection in the partially ionized chromosphere by considering 
the radiative loss and ambipolar diffusion due to the neutral-ion collisions. The reconnection rates (0.01$-$0.03), temperature ($\sim$0.08 MK), and upward outflow velocities 
($\sim$40 km s$^{-1}$) of the plasmoids created by the plasmoid instability correspond well to their characteristic values in chromospheric jets.

Coronal jets are often associated with coronal bright points (CBPs) at their bottom \citep{kri71,gol74,hab81,zhang12a,hong14,ali15}. As one type of long-lived (2$-$48 hr), small-scale 
(10$\arcsec$$-$40$\arcsec$) activities with temperature \citep[1$-$4 MK;][]{kar11} and density \citep[10$^{9}-$10$^{10}$ cm$^{-3}$;][]{uga05} enhancements, CBPs are 
also believed to be heated by magnetic reconnection in the lower corona \citep[e.g.][]{pri94,man96,long98,san07}. Likewise, the intensities of CBPs occasionally show periodic variations 
as a result of repeated and intermittent magnetic reconnections, with the period ranging from a few minutes to $\sim$1 hr \citep{str92,tian08,kar08,zhang12b,ning14,sam15}. 
By performing the potential-field ($\curl\bb=0$) magnetic extrapolation, \citet{zhang12b} found that two neighbouring CBPs on 2007 March 16 were associated with two magnetic 
null points and the corresponding dome-shaped, spine-separatrix topology. Based on the magnetic configurations, they proposed that the CBP evolutions consist of quasi-periodic, 
impulsive flashes and gradual, weak brightening, which are caused by fast null-point reconnection and slow separatrix reconnection, respectively. 
\citet{zhang14c} studied the substructures of a CBP that consisted of two lobes and showed repeated brightenings or flashes on 2009 August 22$-$23. The two lobes  
brightened alternatively and the large-scale overlying coronal loop drifted in the opposite directions during the last two flashes, which is strongly indicative of interchange reconnections.
The double sympathetic events to the east of the CBP with time delay of $<$9 minutes were further studies by \citet{zhang13}. The authors proposed that the likely agent of energy transport 
from the primary CBP to sympathetic events is thermal conduction front. However, the mechanism of jet flow could not be excluded. In this paper, we present observations of two 
groups of jets. The first group of six jets occurred during 
16:30$-$19:00 UT on 2014 September 10, which we call J1$-$J6. All of these jets originated from the same primary CBP, which we call BP1. We will show that the jets flowed along closed 
magnetic loops. We can identify two such loops: one loop connects BP1 with another bright point, which we call BP2; the second loop connects BP1 with a different bright point, which we call 
BP3. The second group of jets took place during 15:55$-$15:58 UT at the boundary of AR 11092 on 2010 August 3. The jets originated from the same primary CBP, which 
we call BP. We could also identify one small-scale loop that connects BP with the AR.
Of particular interest is the multiple blobs in the jets. In Section~\ref{s-data}, we describe the data analysis. Results of the two groups of jets are presented in Section~\ref{s-result}. 
Discussion and summary are arranged in Section~\ref{s-disc} and Section~\ref{s-sum}.

\section{Data analysis} \label{s-data}
The jets on 2014 September 10 were observed by the \textit{Extreme-Ultraviolet Imager} (EUVI) in the \textit{Sun Earth Connection Coronal and Heliospheric Investigation} \citep[SECCHI;][]{how08} 
package of the \textit{Solar TErrestrial RElations Observatory} \citep[STEREO;][]{kai05}. The ahead satellite (hereafter STA) and the behind satellite (hereafter STB) had separation angles of 
$\sim$167$^{\circ}$ and $\sim$161$^{\circ}$ with respect to the Sun-Earth direction. Data from STA were not available at the time of this set of jets, and these jets occurred on the far-side of 
the Sun from SDO.  Therefore, we only observed this set of jets with STB. The four filters of EUVI (171, 195, 284, and 304 {\AA}) have spatial resolution 
of 3.2$\arcsec$ and cadences of 75 s, 300 s, 300 s, and 150 s, respectively. Hence, we mainly used the full-disk 171 {\AA} 
and 304 {\AA} images. Calibration of the EUVI data was performed using the standard Solar Software (SSW) program \textit{secchi\_prep.pro}. The deviation of STB north-south direction 
from the solar rotation axis was corrected. Since the intensity contrast between the jets and background quiet region was very low, we also made base-difference and running-difference 
images to show the jets and CBPs more clearly. The 171 {\AA} and 304 {\AA} images at 17:20:12 UT and 17:19:12 UT before the onset of jets are taken as the base images.

The jets on 2010 August 3 were observed by SDO/AIA in six of the EUV filters (94, 131, 171, 193, 211, 335 {\AA}). Compared with STEREO/EUVI, AIA has much higher spatial resolution 
(1.2$\arcsec$) and time cadence (12 s). The full-disk level\_1 data were calibrated using the standard SSW program \textit{aia\_prep.pro}. The images in various filters were carefully co-aligned 
with accuracy of 0.6$\arcsec$. We performed DEM inversion and studied the temperature properties of the blobs in the jets. The intensity of an optically-thin line $i$ is 
\begin{equation}
I_{i}=\int_{T_{min}}^{T_{max}}\mathrm{DEM}(T)\times R_{i}(T)dT,
\end{equation}
where $\log T_{min}=5.5$ and $\log T_{max}=7.5$ stand for the minimum and maximum temperatures for the integral, $R_{i}(T)$ represents the temperature 
response function of line $i$. The definition and expression of DEM is 
\begin{equation}
\mathrm{DEM}(T)=\frac{d\mathrm{EM}}{dT}=n_{e}^{2}\frac{dh}{dT}, 
\end{equation}
where EM stands for the total column emission measure along the LOS
\begin{equation}
\mathrm{EM}=\int_{T_{min}}^{T_{max}}\mathrm{DEM}(T)dT=\int n_{e}^{2}dh.
\end{equation}
Here, $n_{e}$ denotes the electron number density. 
The DEM-weighted average electron temperature along the LOS is
\begin{equation}
T_{e}=\frac{\int_{T_{min}}^{T_{max}}\mathrm{DEM}\times T\times dT}{\int_{T_{min}}^{T_{max}}\mathrm{DEM}dT}
         =\frac{\int_{T_{min}}^{T_{max}}\mathrm{DEM}\times T\times dT}{\mathrm{EM}}.
\end{equation}
Since the jets were at the AR boundary, background subtraction should be conducted. We took the EUV images at $\sim$15:52 UT before the jets as base images and derived the 
base-difference images during the jets. We carried out DEM reconstructions using the base-difference intensities of the blobs in the six filters and the same program as in \citet{zhang14b}.
To evaluate the uncertainties of the reconstructed DEM curves, 100 Monte Carlo (MC) simulations were conducted for each inversion \citep{cheng12}. The observational parameters are 
summarized in Table~\ref{tbl1}.

\begin{table}
\caption{Description of the observational parameters.}
\label{tbl1}
\tabcolsep 1.5mm
\begin{tabular}{lcccccc}
\hline
Date & Time    & Instrument & $\lambda$ &  Cadence  &   Pixel size & Location \\
        &  (UT)    &                   & ({\AA})       &   (second)  &   (arcsec)   &     \\
\hline
2014/09/10 & 16:30$-$19:00 & STB/EUVI & 171         &  75 & 1.6 & backside \\
2014/09/10 & 16:30$-$19:00 & STB/EUVI & 304         & 150 & 1.6 & backside \\
2014/09/10 & 16:30$-$19:00 & STB/EUVI &195, 284  & 300 & 1.6 & backside \\
2010/08/03 & 15:55$-$15:58 & SDO/AIA   & 94$-$335  & 12  & 0.6 & AR 11092 \\
\hline
\end{tabular}
\end{table}

\section{Results} \label{s-result}
\subsection{Jets on 2014 September 10} \label{s-910}
\begin{figure}
 \centerline{\includegraphics[width=1.0\textwidth,clip=]{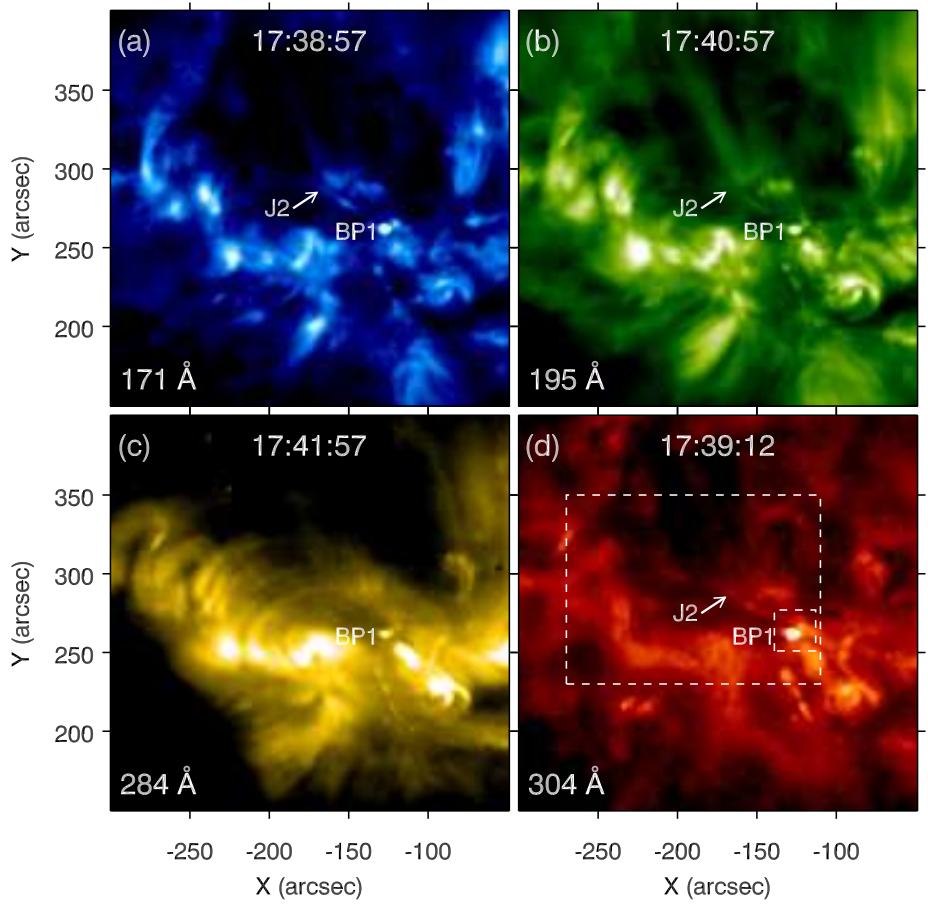}}
 \caption{(a)$-$(d) Four snapshots of EUV images observed by STB/EUVI in 171 {\AA}, 195 {\AA}, 284 {\AA}, and 304 {\AA} around 17:39 UT on 2014 September 10. 
 The homologous jets originate from BP1. The white arrows point to the slim and weak jet (J2) in panels (a), (b), and (d).
 The larger dashed box in panel (d) shows the FOV ($160\arcsec\times120\arcsec$) of Figures~\ref{fig2}$-$\ref{fig7}.}
 \label{fig1}
 \end{figure}

\begin{figure}
 \centerline{\includegraphics[width=1.0\textwidth,clip=]{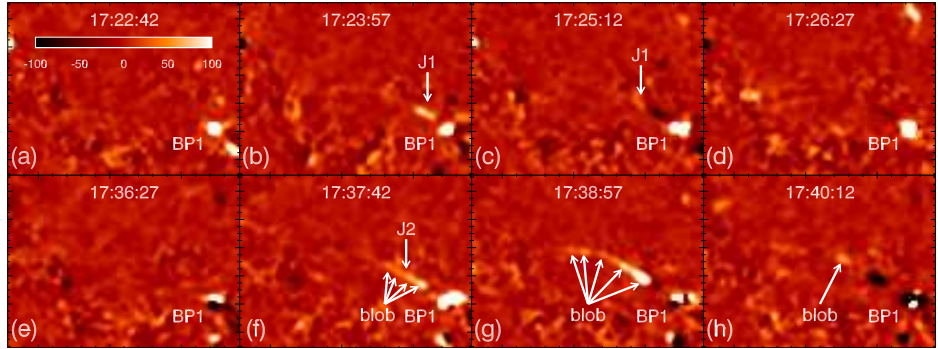}}
 \caption{(a)$-$(h) Eight snapshots of the running-difference images in 171 {\AA} during J1 and J2. In panels (b)$-$(c), the white arrows point to J1. In panels (f)$-$(h), 
 the multiple white arrows point to the bright and compact blobs in J2.}
 \label{fig2}
 \end{figure}

\begin{figure}
 \centerline{\includegraphics[width=1.0\textwidth,clip=]{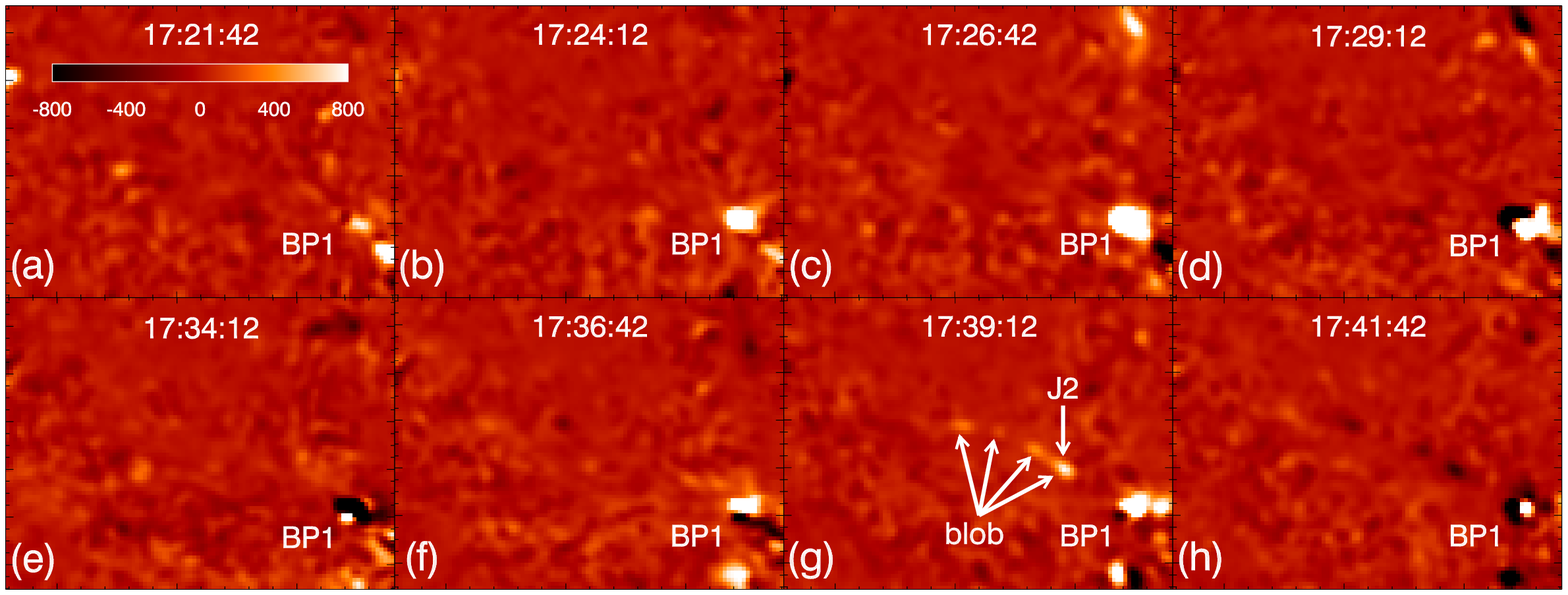}}
 \caption{(a)$-$(h) Eight snapshots of the running-difference images in 304 {\AA} during J1 and J2. In panel (g), the multiple white arrows point to the bright and compact blobs in J2.}
 \label{fig3}
 \end{figure}

\begin{figure}
 \centerline{\includegraphics[width=1.0\textwidth,clip=]{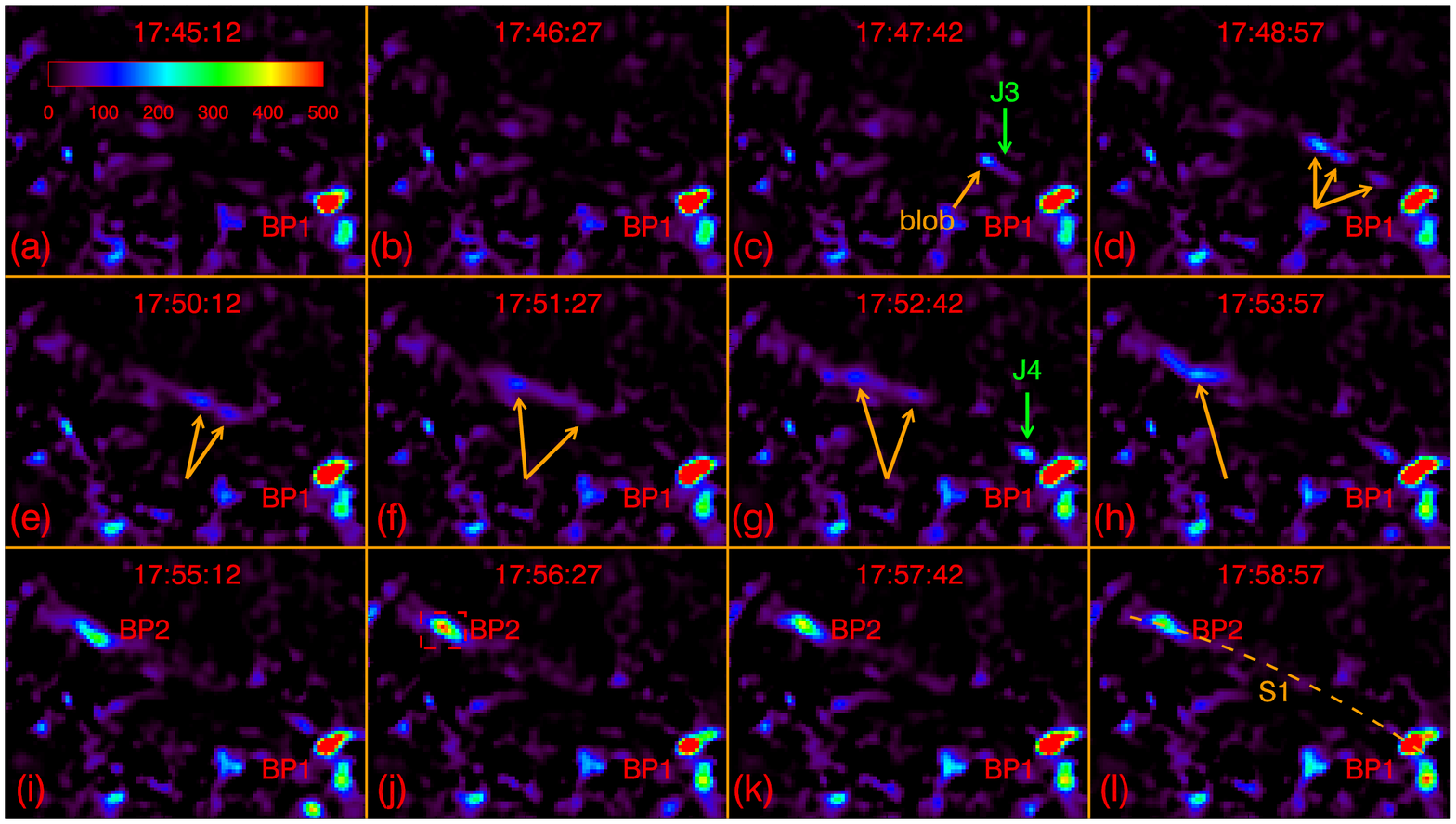}}
 \caption{(a)$-$(l) Twelve snapshots of the base-difference images in 171 {\AA} during J3 and J4. In panels (c)$-$(h), the yellow arrows points to the blobs in J3. In panel (g), the green 
 arrow points to J4. The yellow dashed slice labeled with ``S1'' in panel (l) is used to study the temporal evolutions of the jets and the first sympathetic event (BP2). The length of S1 is 
 149.5$\arcsec$.}
 \label{fig4}
 \end{figure}

\begin{figure}
 \centerline{\includegraphics[width=1.0\textwidth,clip=]{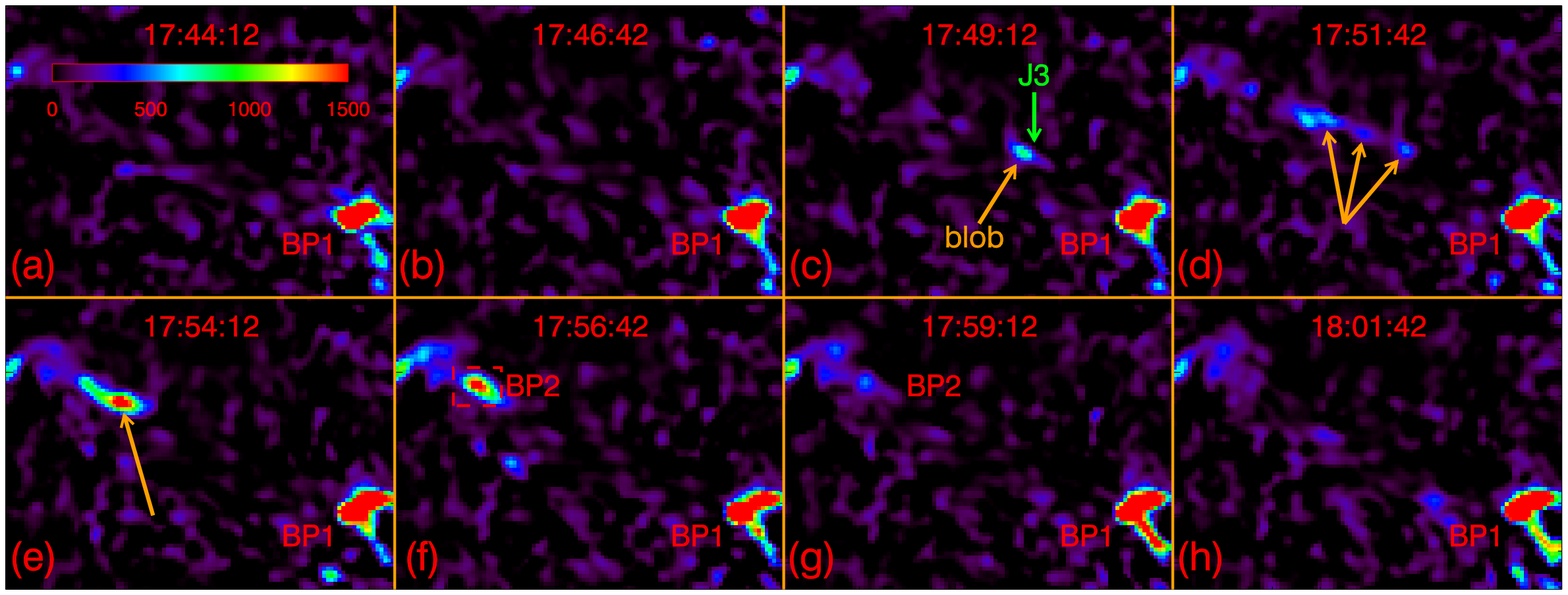}}
 \caption{(a)$-$(h) Eight snapshots of the base-difference images in 304 {\AA} during J3 and J4. In panels (c)$-$(e), the yellow arrows points to the blobs in J3.}
 \label{fig5}
 \end{figure}

\begin{figure}
 \centerline{\includegraphics[width=1.0\textwidth,clip=]{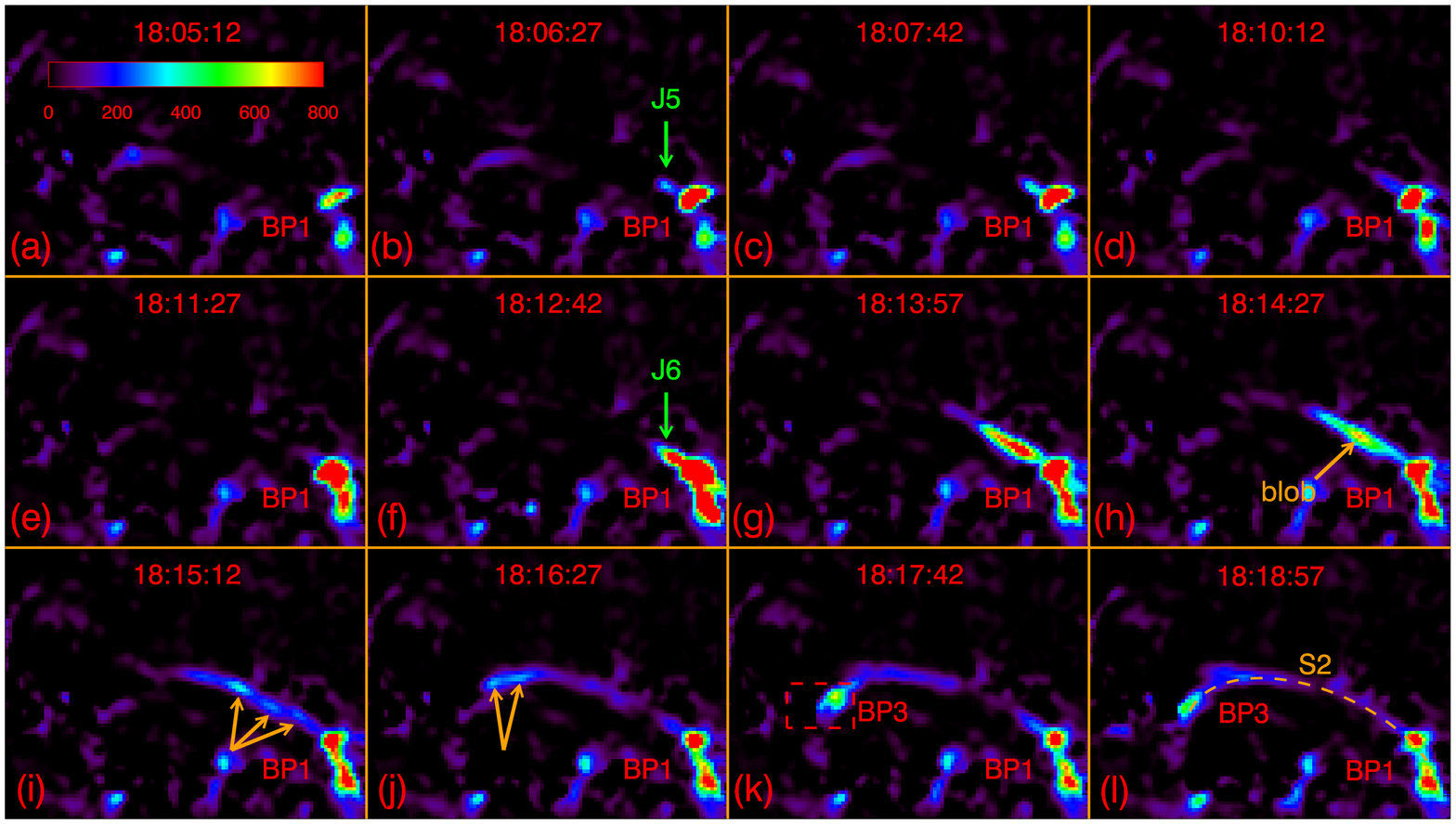}}
 \caption{(a)$-$(l) Twelve snapshots of the base-difference images in 171 {\AA} during J5 and J6. In panel (b), the green arrow points to J5. In panels (h)$-$(j), the yellow arrows points to the 
 blobs in J6 as indicated by the green arrow in panel (f). The yellow dashed slice labeled with ``S2'' in panel (l) is used to study the temporal evolutions of the jets and the second sympathetic 
 event (BP3). The length of S2 is 123.2$\arcsec$.}
 \label{fig6}
 \end{figure}

\begin{figure}
 \centerline{\includegraphics[width=1.0\textwidth,clip=]{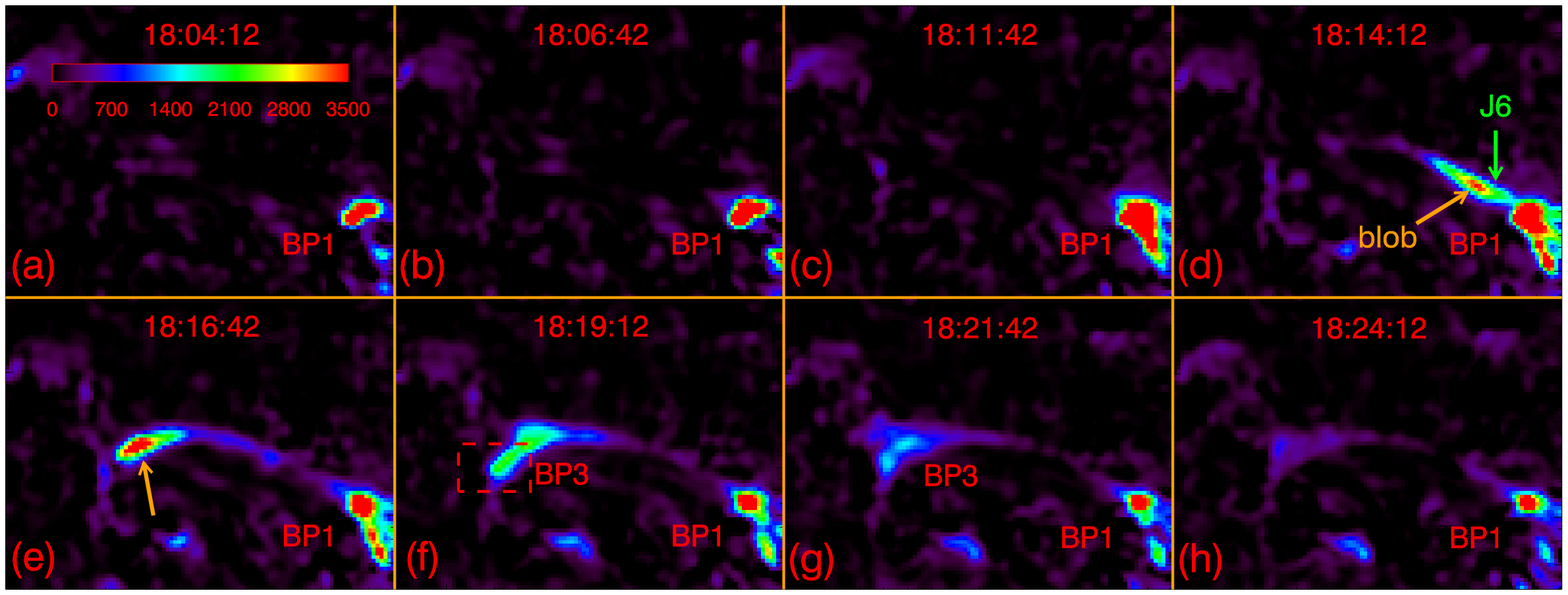}}
 \caption{(a)$-$(h) Eight snapshots of the base-difference images in 304 {\AA} during J5 and J6. In panels (d)$-$(e), the yellow arrows points to the blobs in J6.}
 \label{fig7}
 \end{figure}

Figure~\ref{fig1} shows the EUV images observed in 171 {\AA}, 195 {\AA}, 284 {\AA}, and 304 {\AA} when J2 took place. It is clear that the weak, slim jet was ejected in the northeast 
direction from BP1 located at (-130$\arcsec$, 260$\arcsec$) in the quiet region. Due to the impulsive nature of the jet and the 
low cadence of 284 {\AA}, the jet was absent in 284 {\AA} with formation temperature of $\sim$2 MK. Nevertheless, bright coronal loops are evident in 284 {\AA}.
The rectangular dashed box with size of $160\arcsec\times120\arcsec$ in panel (d) represents the field-of-view (FOV) of the panels in Figures~\ref{fig2}$-$\ref{fig7}.

Figure~\ref{fig2} shows eight snapshots of the 171 {\AA} running-difference images during J1 and J2. With the brightness of BP1 increasing, the jet (J1) was ejected outwards from BP1 in the 
northeast direction (see panel (b)). After $\sim$17:36 UT, J2 occurred at the same place. Interestingly, we find bright and compact features in J2. Four and five such features are identified 
by eye at 17:37:42 UT and 17:38:57 UT as pointed by the white arrows. In view of the extraordinary resemblance to the blobs reported by \citet{zhang14b}, we also take the features as blobs.
The discrete, circular or elliptical blobs with sizes of 4.5$-$7.5 Mm moved along the jets, like sliding pearls along a necklace.

The 304 {\AA} running-difference images during J1 and J2 are displayed in Figure~\ref{fig3}. Due to the short lifetime of J1 and the low cadence of 304 {\AA}, J1 was hardly identified. 
However, multiple blobs are distinct in 304 {\AA} at 17:39:12 UT, as pointed by the white arrows, which is consistent with those in Figure~\ref{fig2}(g).
 
Figure~\ref{fig4} shows the 171 {\AA} base-difference images during J3 and J4. Starting from $\sim$17:47 UT, the jet (J3) propagated in the northeast direction from BP1 and terminated at 
BP2 about 9 minutes later (see panel (j)). BP2 is located at (-237$\arcsec$, 314$\arcsec$), which is $\sim$120$\arcsec$ away from BP1. Afterwards, the brightness 
of BP2 decreased gradually until it disappeared. Like J2, J3 is not coherent. It consists of blobs propagating along the jet, as pointed by the yellow arrows in panels (c)$-$(h). The sizes of 
the blobs range from 6.5 Mm to 8.0 Mm. The short J4 appeared at $\sim$17:52 UT and disappeared at $\sim$17:57 UT. The 304 {\AA} base-difference images during J3 and J4 are presented 
in Figure~\ref{fig5}. Similar to Figure~\ref{fig4}, the blobs propagated along J3 and stopped at BP2, as indicated by the yellow arrows in panels (c)$-$(e).
 
Figure~\ref{fig6} shows the 171 {\AA} base-difference images during J5 and J6. The short jet (J5) was ejected out of BP1 in the northeast direction after $\sim$18:06 UT, as pointed by the 
green arrows in panel (b). After $\sim$18:12 UT, J6 spurted from BP1 in the same direction and propagated to BP3, which is located at (-230$\arcsec$, 280$\arcsec$) 
and is $\sim$105$\arcsec$ away from BP1. Afterwards, the intensity of BP3 decreased gradually and finally it disappeared. Note that the trajectory of J6 is close to that of J3 in the 
initial phase, but different at the ending phase, though both J3 and J6 produced remote brightening at the other end of large-scale coronal loops. Likewise, multiple blobs are recognized 
in J6 by eye, as indicated by the yellow arrows in panels (h)$-$(j). Figure~\ref{fig7} shows the 304 {\AA} base-difference images during J5 and J6. J5 is not obvious in 304 {\AA} due to the low 
cadence. However, J6 is very clear with blobs at 18:14:12 UT and 18:16:42 UT. Table~\ref{tbl2} summarizes the informations of the homologous jets, including the wavelengths of 
observation ($\lambda$), begin times in 171 {\AA}, end times in 171 {\AA}, lifetimes, apparent lengths, apparent velocities, and connected CBP.

\begin{table}
\caption{Informations of the homologous coronal jets (J1$-$J6) on 2014 September 10.}
\label{tbl2}
\tabcolsep 2.0mm
\begin{tabular}{cccccccc}
\hline
Jet & $\lambda$ & $t_{begin}$ & $t_{end}$ & Lifetime & Length   & Velocity         & CBP \\
      &  ({\AA})      & (UT)            & (UT)         & (s) & (arcsec) & (km s$^{-1}$) &        \\
\hline
J1 & 171         & 17:23:57  & 17:26:27 & 150$\pm$75 &   26.2$\pm$1.6 & 145$\pm$15 & - \\
J2 & 171, 304 & 17:37:42  & 17:41:27 & 225$\pm$75 &   50.7$\pm$1.6 & 203$\pm$8 & - \\
J3 & 171, 304 & 17:47:42  & 17:53:57 & 375$\pm$75 & 124.6$\pm$1.6 & 381$\pm$18 & BP2 \\
J4 & 171         & 17:52:42  & 17:56:27 & 225$\pm$75 &   28.8$\pm$1.6 & 175$\pm$13 & - \\
J5 & 171         & 18:06:27  & 18:11:27 & 300$\pm$75 &   16.7$\pm$1.6 & 222$\pm$34 & - \\
J6 & 171, 304 & 18:12:42  & 18:18:57 & 375$\pm$75 & 108.2$\pm$1.6 & 249$\pm$3 & BP3 \\                
\hline
\end{tabular}
\end{table}
 
 \begin{figure}
 \centerline{\includegraphics[width=1.0\textwidth,clip=]{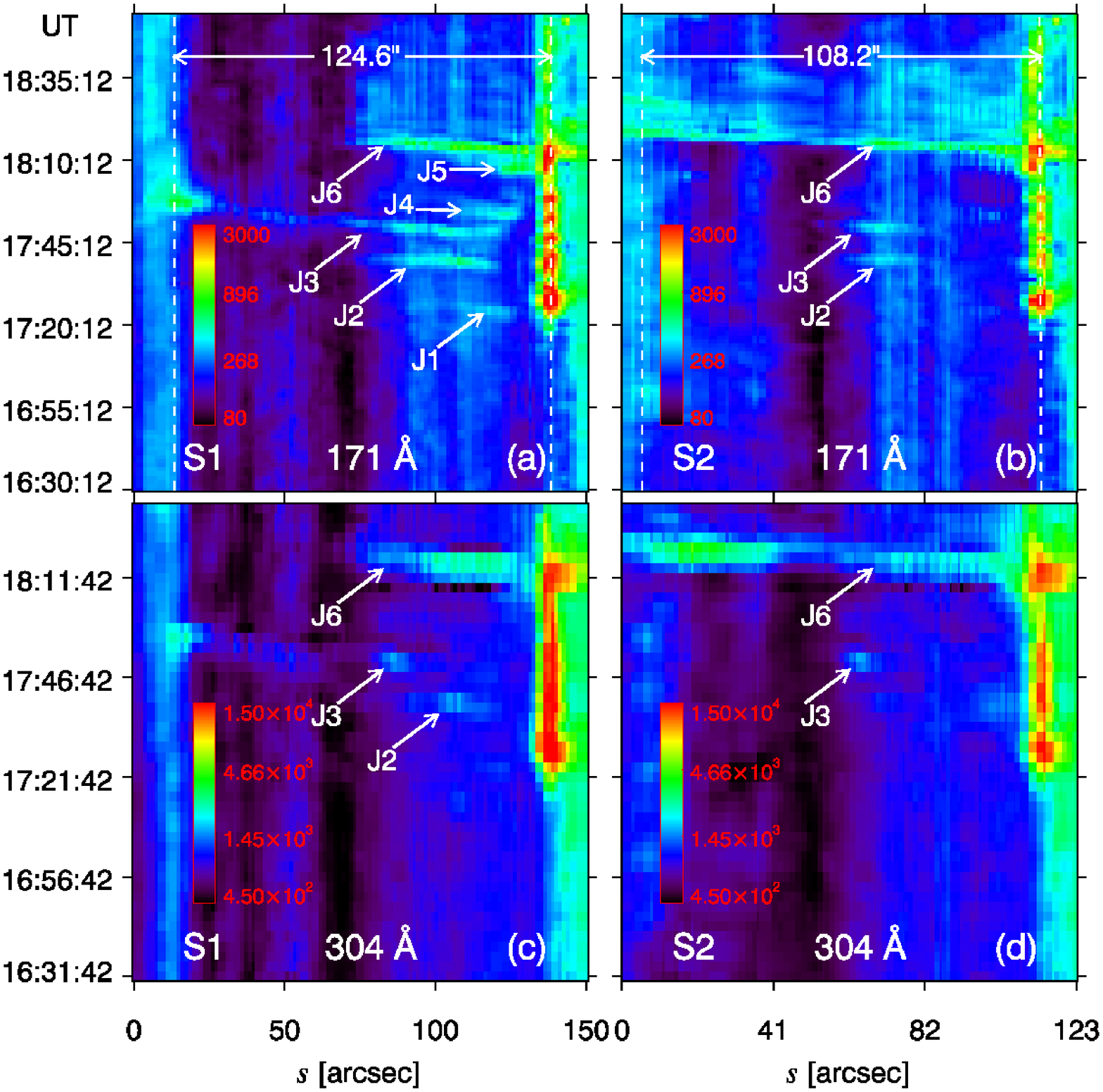}}
 \caption{Time-slice diagrams of S1 (left panels) and S2 (right panels) in 171 {\AA} (top panels) and 304 {\AA} (bottom panels). The homologous jets (J1$-$J6) are bright and inclined 
 structures whose inverses of slope stand for the apparent velocities of the jets, as pointed by the white arrows. The distances between the vertical dashed lines represent the distances 
 between BP1 and BP2 (BP3) in the upper panels.}
 \label{fig8}
 \end{figure}
 
In order to investigate the temporal evolution of the jets, we extract two curved slices. The first slice (S1) is superposed on Figure~\ref{fig4}(l) with yellow dashed line, which passes 
through BP1 and BP2. The second slice (S2) is superposed on Figure~\ref{fig6}(l) with yellow dashed line, which passes through BP1 and BP3. The time-slice diagrams of S1 and S2
in 171 {\AA} and 304 {\AA} are illustrated in Figure~\ref{fig8}. The $x-$ and $y-$axes denote the distances from the left endpoints of the slices and the time (UT) in each panel. In panel 
(a), the homologous jets (J1$-$J6) are represented by the the bright and inclined structures whose inverses of slope stand for their apparent speeds, being 145$\pm$15, 203$\pm$8, 
381$\pm$18, 175$\pm$13, 222$\pm$34, and 224$\pm$7 km s$^{-1}$, respectively. It is clear that J3 propagates along S1 and reaches the left endpoint of S1, producing the sympathetic 
CBP, i.e., BP2 around 17:56 UT. In panel (b), only 
J2, J3, and J6 whose maximum apparent lengths are longer than J1, J4, and J5 are evident. J6 propagates along S2 and reaches the left endpoint of S2, producing the sympathetic CBP, 
i.e., BP3 around 18:17 UT. The more accurate apparent speed of J6 derived from panel (b) is 249$\pm$3 km s$^{-1}$. Owing to the low cadence of the 304 {\AA} filter, only some 
of the jets (J2, J3, and J6) are clearly identified (see panels (c)$-$(d)). The intervals of the recurrent jets range from 5 to 15 minutes.

 \begin{figure}
 \centerline{\includegraphics[width=0.9\textwidth,clip=]{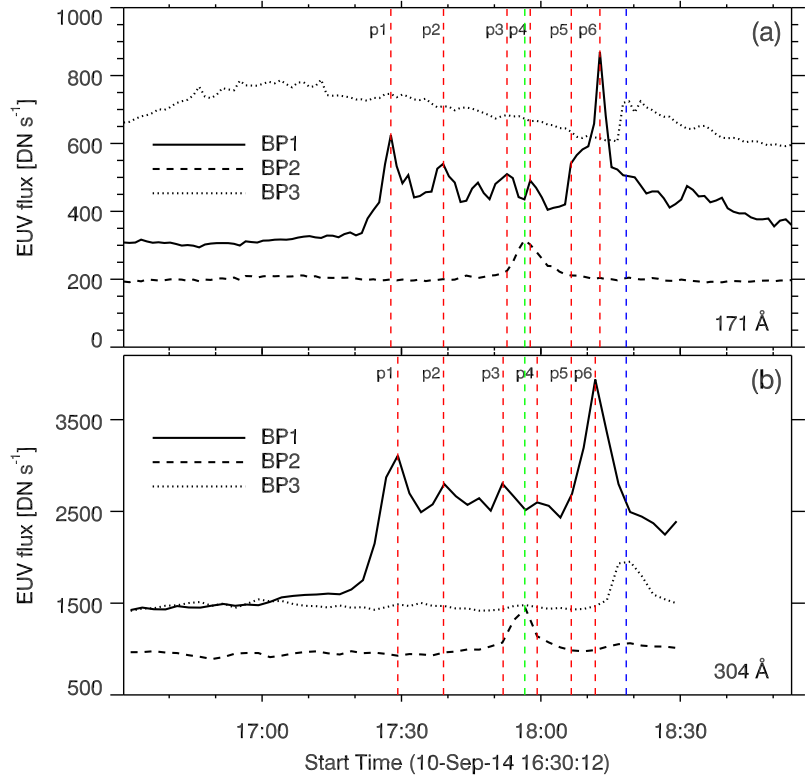}}
 \caption{(a)$-$(b) Light curves of BP1 (solid lines), BP2 (dashed lines), and BP3 (dotted lines) in 171 {\AA} and 304 {\AA}. The light curves are calculated to be the average intensities 
 of BP1, BP2, and BP3 within the small dashed boxes of Figure~\ref{fig1}(d), Figure~\ref{fig4}(j), and Figure~\ref{fig6}(k), respectively. The vertical red, green, and blue lines represent 
 the peak times of BP1 (p1$-$p6), BP2, and BP3.}
 \label{fig9}
 \end{figure}
 
In Figure~\ref{fig9}, we plot the light curves of BP1, BP2, and BP3 in 171 {\AA} and 304 {\AA} with solid, dashed, and dotted lines, respectively. The light curves are calculated to be the 
average intensities of BP1, BP2, and BP3 in the small box of Figure~\ref{fig1}(d), Figure~\ref{fig4}(j), and Figure~\ref{fig6}(k). During 17:20$-$18:30 UT, BP1 experiences several 
flashes, whose peak times (p1$-$p6) are labeled with red dashed line. The last flash is the strongest, which is similar to the case of CBP on 2009 August 22$-$23 observed in 
SXR \citep{zhang14c}. The intensities of BP2 increase slowly from 17:40 UT and rapidly from $\sim$17:50 UT before reaching the maxima at $\sim$17:56:30 UT. Afterwards, they decrease 
to the initial levels at $\sim$18:10 UT. The lifetime of BP2 is 27.5$\pm$2.5 minutes. Since BP1 reaches the third peak (p3) at $\sim$17:52:30 UT, the time delay between BP2 and 
BP1 is 240$\pm$75 s. The intensities of BP3 increase from $\sim$18:13:30 UT and reach the maxima at $\sim$18:17:30 UT before declining to the initial levels at $\sim$18:43 UT. The 
lifetime of BP3 is 27.4$\pm$2.5 minutes. Considering that the peak time of the sixth flash of BP1, i.e. p6, is $\sim$18:12:30 UT, the time delay between BP3 and BP1 is 300$\pm$75 s.

\subsection{Jets on 2010 August 3} \label{s-803}

\begin{figure}
 \centerline{\includegraphics[width=1.0\textwidth,clip=]{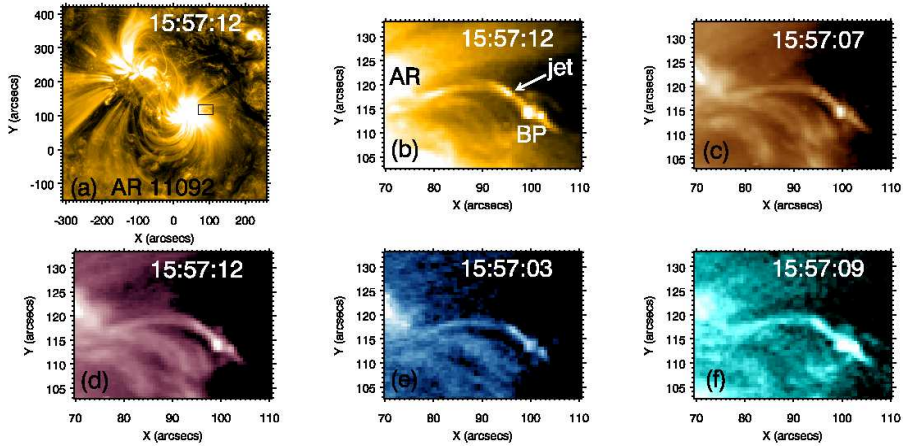}}
 \caption{(a) EUV image of the AR 11092 at 15:57:12 UT on 2010 August 3 observed by SDO/AIA in 171 {\AA}. 
 (b)$-$(f) Amplified EUV images of the jets and BP at the AR boundary observed in five of the AIA 
 filters, i.e., 171 {\AA}, 193 {\AA}, 211 {\AA}, 335 {\AA}, and 131 {\AA}. 
 The FOV of these images is indicated by the small black rectangular box in panel (a).}
 \label{fig10}
 \end{figure}

In Figure~\ref{fig10}, the 171 {\AA} image of the AR 11092 at 15:57:12 UT is shown in panel (a). The jets were located at the western AR boundary in the black rectangular box 
($40\arcsec\times30\arcsec$). In panels (b)-(f), we plot the EUV images of the jets observed in 171, 193, 211, 335, and 131 {\AA} around 15:57 UT. It is clear that the jets originated from the 
bright point (BP) and propagated in the northeast direction along the small-scale closed loop that is $\sim$28 Mm in length before flowing into the AR. The bright jets with enhanced emission 
measure were observed in all the EUV filters, though we show the jets only in five filters.

\begin{figure}
 \centerline{\includegraphics[width=1.0\textwidth,clip=]{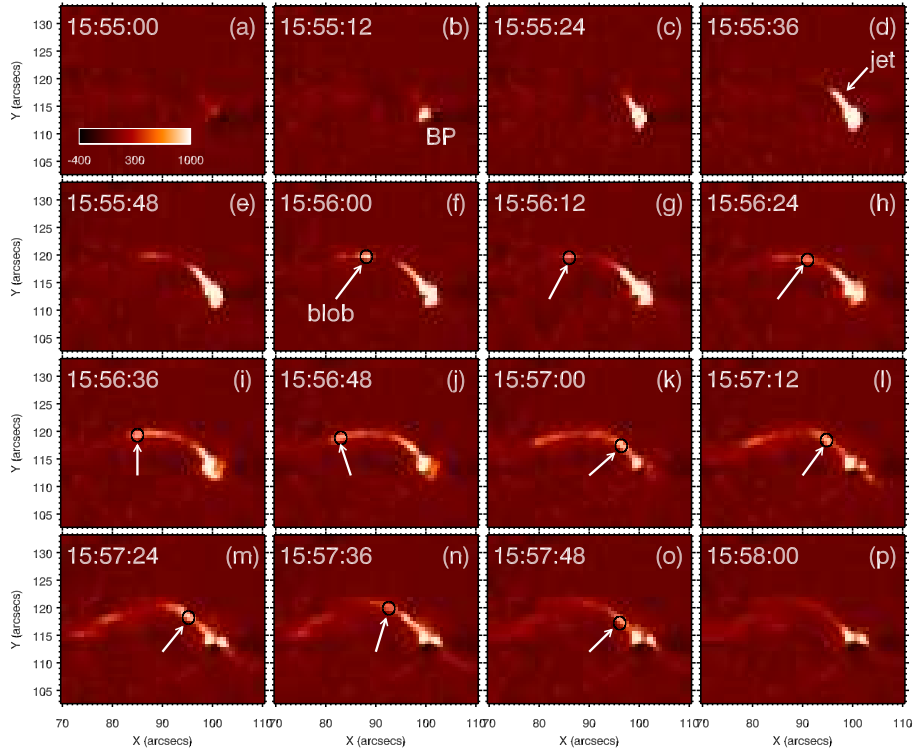}}
 \caption{(a)$-$(p) Base-difference 171 {\AA} images of the jets and BP. In panels (f)$-$(o), the white arrows point to the blobs 
 encircled by the black circles.}
 \label{fig11}
 \end{figure}

Compared with the first group of jets, the second group was short-lived. The whole evolution is displayed in the 171 {\AA} base-difference images of Figure~\ref{fig11}. 
The jet started at $\sim$15:55 UT and moved in the northeast direction, with the intensity of BP at the bottom increasing. Interestingly, the jet was not coherent and smooth. Instead, 
there were bright and compact structures similar to the blobs in the jets observed by EUVI. In panels (f)-(o), we encircle the blobs with black circles that are 2.4$\arcsec$ in diameter. 
The blob appeared at $\sim$15:56:00 UT (panel (f)) and propagated along the closed loop for $\le$12 s. A new blob appeared at 15:56:24 UT (panel (h)) and lasted until $\sim$15:56:48 UT 
(panel (j)). The velocity of this blob is calculated to be 238 km s$^{-1}$. Then, another blob appeared close to the BP and flowed forward until $\sim$15:57:12 UT (panel (l)). 
Afterwards, three blobs were identified during the late phase of evolution. As the brightness of BP decreased slowly, the jet faded and disappeared. Since the jets flowed into the AR, 
sympathetic CBPs were not observed at the remote footpoint of the loop.
 
\begin{figure}
 \centerline{\includegraphics[width=1.0\textwidth,clip=]{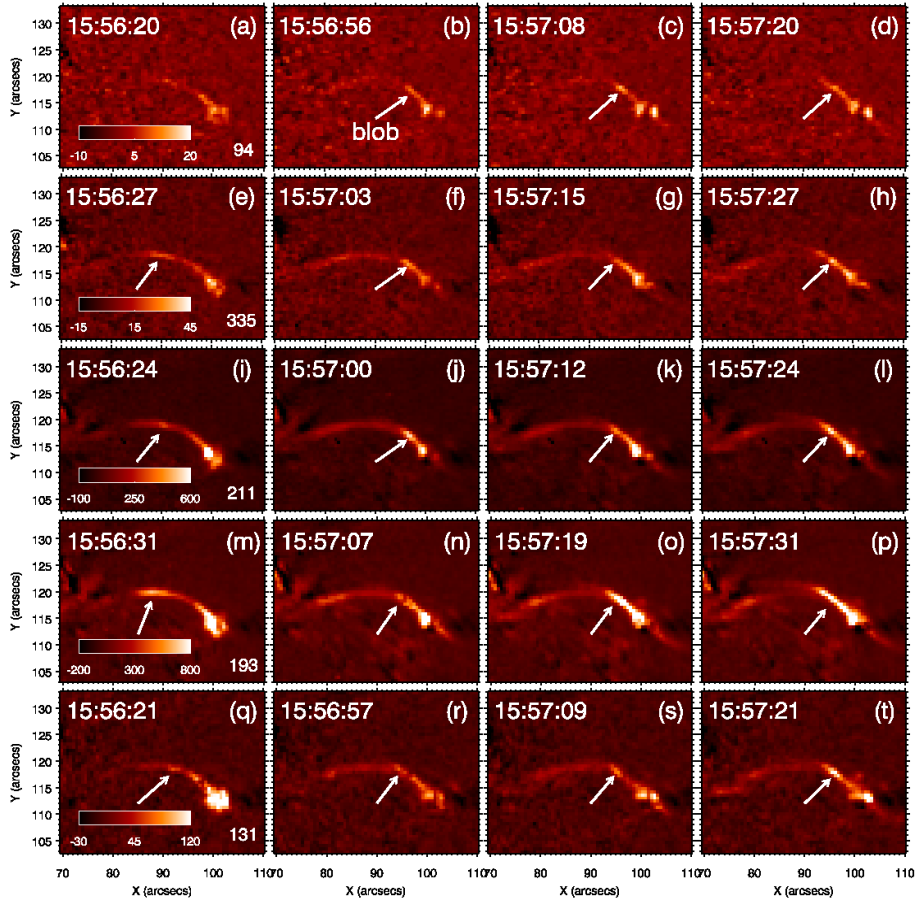}}
 \caption{From top to bottom row: Base-difference EUV images of the jets and BP in 94, 335, 211, 193, and 131 {\AA}. The white arrows point to the blobs.}
 \label{fig12}
 \end{figure}

The blobs were visible not only in 171 {\AA}, but also in the other wavelengths. In Figure~\ref{fig12}, the base-difference images in 94, 335, 211, 193, and 131 {\AA} are shown from 
top to bottom row, with the blobs being indicated by the white arrows. In each column, the observing times are very close. As in 171 {\AA}, the blobs are bright and compact features in 
the closed loop. It should be noted that a blob is not always evident and striking in all the wavelengths. The presence of blobs in multiple wavelengths suggests their multithermal nature.
 
\begin{figure}
 \centerline{\includegraphics[width=1.0\textwidth,clip=]{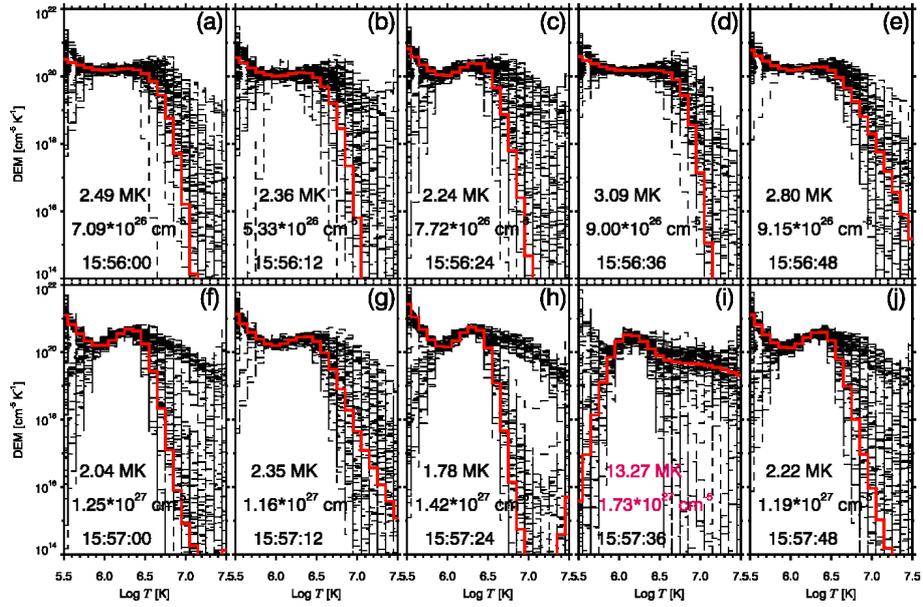}}
 \caption{(a)$-$(j) DEM profiles of the blobs as indicated by the arrows in Figure~\ref{fig11}(f)-(o). 
 The red solid lines stand for the best-fitted DEM curves from the observed values. 
 The black dashed lines represent the reconstructed curves from the 100 MC simulations. 
 The corresponding $T_e$ (MK) and EM (cm$^{-5}$) are displayed.}
 \label{fig13}
 \end{figure}

In order to study the temperature properties of the blobs, we performed DEM reconstructions as described in Sect.~\ref{s-data}. Figure~\ref{fig13} shows the DEM profiles of the 
ten blobs as indicated by the arrows in Figure~\ref{fig11}(f)-(o). The red solid lines stand for the best-fitted DEM curves from the observed values, while the black dashed lines represent the 
reconstructed curves from the 100 MC simulations. Except the ninth one, all the profiles have similar shapes in the range of $5.5<\log T[K]<7.5$, with most of the emissions coming from the 
low-temperature plasma. The uncertainties of the DEM curves are minimum in the range of $5.8<\log T[K]<6.5$ and increase significantly towards the low-temperature and high-temperature 
ends. The calculated EM of the blobs ranges from 5$\times$10$^{26}$ cm$^{-5}$ to 1.4$\times$10$^{27}$ cm$^{-5}$. Assuming that the LOS depths of the blobs equal to the widths, i.e., 
2.4$\arcsec$, the electron number densities of the blobs are estimated to be (1.7$-$2.8)$\times10^9$ cm$^{-3}$. The calculated $T_e$ of the blobs ranges from 1.8 MK to 3.1 MK except the 
ninth one, and most of them are between 2 MK and 3 MK.
 
\section{Discussions} \label{s-disc}
So far, studies of plasmoids or blobs in coronal jets are few. Using the multiwavelength observations of AIA on 2011 July 22, \cite{zhang14b} studied the recurring blobs. 
In this paper, we provide additional evidences of multiple, recurring blobs in the homologous jets. The first group was observed by STB/EUVI and the second group was observed by AIA.
In Table~\ref{tbl3}, we compare the parameters of the blobs. On one hand, the measured sizes of the blobs observed by EUVI are 2$-$5 times larger than those observed by AIA. 
Considering that the resolution of EUVI is lower than that of AIA, the blobs observed by EUVI may not be fully resolved and the obtained blob sizes with large uncertainties may not be 
reliable. On the other hand, both the sizes of blobs and CBP at the bottom on 2010 August 3 are about two times smaller than those for the event on 2011 July 22. In chromospheric 
anemone jets, the sizes of blobs (0.3$-$1.5 Mm) as well as the base loops are even smaller \citep{sin12}, implying that the same process may exist in different scales in the solar atmosphere. 
The lifetime of the blobs in coronal jets observed by AIA ranges from 12 s to 60 s, which is similar to the typical value of chromospheric blobs \citep{sin12}. The apparent speeds of the blobs 
in the order of coronal Alfv\'{e}n speed, i.e., hundreds of km s$^{-1}$, are consistent in the FOVs of AIA and EUVI. For the events observed by AIA, the DEM-weighted average temperatures 
of the blobs agree with each other, suggesting that the method of inversion we use is correct. Besides, the DEM profiles are similar. The possible causes of problematic result of inversion for 
the ninth blob in Figure~\ref{fig13}(i) and large uncertainties of all the profiles at the low-temperature and high-temperature ends are as follows. 1) The intensities of the blobs are not strong 
enough, i.e, the signal-to-noise ratios are not high enough. 2) The observing times of the six filters for inversion are not exactly the same. 
Therefore, the positions of the blobs may be misplaced more or less even though we take the AIA images at the nearest times. Assuming that the velocity of a blob is 200 km s$^{-1}$, 
the maximum displacement of the blob in the EUV images is 1.6 Mm, which is close to the size of blobs. In this regard, multiwavelength observations with much higher cadence are required 
in the future. The rough estimations of the number density of blobs are also in the same range on the assumption that the LOS depth equals to the width or size. Despite that the jets along 
open magnetic fields are prevailing, reports on the observations of coronal jets along closed loops are few \citep{zhang13,zhang14c}. The blobs in jets associated with both open and closed 
magnetic fields indicate that this kind of bright and compact structures are ubiquitous. According to the previous theoretical and numerical studies \citep[e.g.,][]{fur63,dra06,bar08}, the blobs 
are most probably plasmoids created by the TMI of the CS where magnetic reconnection takes place.

\begin{table}
\caption{Comparison of the parameters of the blobs.}
\label{tbl3}
\tabcolsep 1.5mm
\begin{tabular}{lccccccc}
\hline
Date & Instr. & Size   &  Lifetime   & Velocity          & $T_e$ & $n_e$ & Mag. Field  \\
        &           & (Mm) &  (second)  & (km s$^{-1}$) &  (MK)  & (10$^9$ cm$^{-3}$) &  \\
\hline
2011/07/22\tabnote{Event studied by Zhang \& Ji (2014b)} & AIA   &  $\sim$3     & 24$-$60      & 120$-$450 & 0.5$-$4.0 & 1.5$-3.5$ & open  \\
2010/08/03 & AIA   &  $\sim$1.7  & 24$\pm$12 & $\sim$238   & 1.8$-$3.1 & 1.7$-$2.8 & closed   \\
2014/09/10 & EUVI &  4.5$-$9.0  & -                 & 140$-$380  &    -            & -               & closed    \\
\hline
\end{tabular}
\end{table}

Sympathetic phenomena are common in the solar atmosphere because of the complexity and interconnection of the magnetic field lines. Sympathetic flares have been extensively reported and 
investigated in the past decades \citep[e.g.][]{han96,lang89,mas09,wang12,deng13,sun13,yang14,liu15a}. The possible agents of energy transported from the source region of energy release to the 
remote footpoints are nonthermal electrons \citep{tang82,naka85,mart88}, thermal conduction front \citep{rust85,bast92}, shock waves \citep{mach88}, and reconnection outflows \citep{han96,nis97}. 
In a cartoon model, \citet{mach88} compared the velocities of different agents, among which nonthermal electrons have the fastest speed (about one third of the speed of light). The 
electrons are followed by thermal conduction, plasma shock, and evaporated material. Therefore, the time delays between the start/peak times of the main and sympathetic events are different 
for different energy agents. In the rare case of double sympathetic events studied by \citet{zhang13}, the time delays between BP1 and BP2 (BP3) are less than 9 minutes. 
The authors proposed that thermal conduction plays a role in the energy transport. Nevertheless, it is not easy to distinguish thermal conduction and jet flow in observation. In this study, the jets 
(J3 and J6) reached the remote footpoints of the pre-existing, large-scale coronal loops and produced sympathetic CBPs (BP2 and BP3) observed in 171 {\AA} and 304 {\AA}. The pre-existing 
loops may have lower electron number density and emission measure so that they are not clearly revealed in 284 {\AA} before the jets. The time delays between BP1 and BP2 (BP3) are 
approximately 240$\pm$75 s (300$\pm$75 s). Considering that the lengths of the coronal loops between BP1 and BP2 (BP3) are $\sim$124.6$\arcsec$ and $\sim$108.2$\arcsec$ as indicated 
in the upper panels of Figure~\ref{fig8}, the velocities of the energy transport are estimated to be 376.4$_{-90}^{+171}$ and 261.5$_{-52}^{+87}$ km s$^{-1}$, which are consistent with 
the apparent velocities of J3 and J6. On one hand, the transit times of the nonthermal electrons from BP1 to BP2 and BP3 are $\sim$1 s, which are significantly shorter than the observed 
time delays. On the other hand, thermal conduction timescales of the coronal loops are $\tau_{c}=4\times10^{-10}n_{e}L^2T_{e}^{-5/2}$, where $n_e$, $T_e$, and $L$ represent the number density 
of the jets, temperature of the jets, and length of the loops \citep{car94}. Assuming that $n_e=10^9$ cm$^{-3}$ and $T_e=2$ MK, the values of $\tau_{c}$ between BP1 and BP2 (BP3) 
are estimated to be $\sim$5770 s ($\sim$4351 s), which are much longer than the observed values. If the thermal conduction front really plays a dominant role, the temperatures of BP1 and jets 
should be 6$-$7 MK in the case of $n_e=10^9$ cm$^{-3}$, which seems to be unlikely because the response of the 171 {\AA} filter decreases from the maximum value by 3$-$4 orders of 
magnitude at such high temperature. Interestingly, we identified multiple and recurring blobs in the jet flows, which is in favor of the jet flow rather than thermal conduction front to be the most 
probable energy agent because conduction front should result in smooth brightenings rather than clumpy brightenings.
\section{Summary} \label{s-sum}

In this paper, we report our multiwavelength observations of two groups of homologous jets. The first was observed by STB/EUVI on 2014 September 10 and the second was observed by 
SDO/AIA on 2010 August 3. The main results are summarized as follows:

\begin{enumerate}
\item The first group of recurring jets originated from BP1 and propagated in the northeast direction along large-scale, closed coronal loops during the six eruptions (J1$-$J6). 
The intervals of the recurrent jets range from 5 to 15 minutes. Two of the jets (J3 and J6) produced sympathetic CBPs (BP2 and BP3) after reaching the remote footpoints of 
the loops. The peak times of the sympathetic CBPs were delayed by 240$-$300 s relative to BP1.
\item The jets were not coherent. Instead, they were composed of bright and compact structures, i.e., blobs. The sizes and apparent velocities of the blobs are 4.5$-$9 Mm and 
140$-$380 km s$^{-1}$, respectively. The existence of multiple blobs in the jets suggests that the sympathetic CBPs are caused by jet flows rather than thermal conduction front. 
\item The second group of jets originated from BP and propagated in the northeast direction along a small-scale, closed coronal loop at the boundary of AR 11092. 
Sympathetic CBPs were not observed at the remote footpoint of the loop (although it is possible that the remote footpoint was hidden in the glare of the AR). 
Like the first group, we also identified blobs in the jets. The sizes and apparent velocities of the blobs are $\sim$1.7 Mm and $\sim$238 km s$^{-1}$, respectively.
\item Using the AIA base-difference EUV images, we performed DEM inversions and derived the DEM profiles of the blobs. The blobs are multithermal with temperatures of 
1.8$-$3.1 MK. The estimated number density of the blobs are (1.7$-$2.8)$\times$10$^9$ cm$^{-3}$. Additional case studies with high resolution and cadence are expected in the future.
Multidimensional MHD simulations are underway to investigate the nature of plasmoids and mechanism of sympathetic CBPs.
\end{enumerate}

\begin{acks}
The authors acknowledge the referee for detailed comments and valuable suggestions. We are also grateful to 
E. Pariat, L. Ni, T. Li, Y. H. Yan, S. L. Ma, P. F. Wyper, C. R. DeVore, and P. Syntelis for constructive discussions. STEREO/SECCHI data are provided 
by a consortium of US, UK, Germany, Belgium, and France. SDO is a mission of NASA\rq{}s Living With a Star Program. AIA data are courtesy of the NASA/SDO science 
teams. QMZ is supported by NSFC No. 11303101, 11333009, and 11473071. H. S. Ji is supported by the Strategic Priority Research Program$-$The Emergence of Cosmological Structures of 
the CAS, Grant No. XDB09000000. Yingna Su is supported by NSFC 11473071, Youth Fund of Jiangsu BK20141043, and the One Hundred Talent Program of Chinese Academy 
of Sciences.
\end{acks}

\end{article} 

\end{document}